\documentclass{article}[12pt]
\usepackage{amssymb}
\usepackage{amsmath}
\usepackage{graphicx}
\usepackage{setspace}
\usepackage[utf8]{inputenc}
\begin{document}
\title{Yukawa coupling and one loop inflation in the light of CMB}
\date{}
\maketitle
\vspace{-15mm}
\begin{center}
\large{Anupama B\footnote{21phph19@uohyd.ac.in} and P K Suresh\footnote{sureshpk@uohyd.ac.in}\\}
\small{School of Physics, University of Hyderabad, \\Hyderabad - 500046. India.\\Corresponding authors: $^1$21phph19@uohyd.ac.in, $^2$sureshpk@uohyd.ac.in}
\end{center}
\begin{abstract}
The one loop inflation stemming from the superstring theory and associated Yukawa coupling arising from supersymmetric interactions is examined with CMB. The Yukawa coupling can exist beyond standard model particle physics sector. The tensor-to-scalar ratio of the loop inflation is found consistent with the recent CMB results for the Yukawa coupling from cosmology. The alternative constraint on the Yukawa coupling from loop inflation may play a crucial role in validating inflationary model originating from supersymmetry and string theory. The outcomes of the study may be helpful in the phenomenological realisation of string theory.
\end{abstract}


\section{Introduction}
\label{sec:intro}
Inflation is considered a period of exponential expansion in the evolution history of the early universe \cite{guth}. It stretches the zero-point quantum oscillations of a homogeneous scalar field called the inflaton, responsible for inflation. One of the important features of inflation includes the generation of scalar and tensor perturbations that respectively lead to the formation of large scale structure of the universe and the generation of primordial gravitational waves. The fractional contribution of these tensor and scalar modes can be quantified using the ratio of their amplitudes, known as the tensor-to-scalar ratio $(r)$. In standard parlance, inflationary models use a single scalar field to describe the dynamics of inflation, but over the years, many single and multi-field inflationary models have also been proposed \cite{enc}. The validation of these models depends primarily on the tensor-to-scalar ratio. The tensor-to-scalar ratio of most of these models is found incompatible with the estimates from various cosmological observations \cite{6, 7, 8, 9, 10, planck, 12}. Therefore, it is interesting to explore inflationary models that can possibly make the tensor-to-scalar ratio compatible with the current \cite{planck} and future projected estimates \cite{Lite, s4}.\\
\\
Apart from the usual classical models, inflation can be described using supersymmetric string theory that predicts a 10-dimensional spacetime with gauge bosons and fermions \cite{all,review}. There are five types of superstring theories, such as type-I, type-IIA, type-IIB, heterotic $SO(32)$ and heterotic $E_8 \times E_8$.  Strings can be considered as 1-dimensional analogues of higher dimensional dynamical objects called $p$-branes. Open strings can be localised to higher dimensional membranes called D-branes by imposing Dirichlet boundary conditions to their endpoints, whereas closed strings enjoy the freedom to propagate through extra dimensions called the bulk \cite{WS,jp}.  Compactification of extra dimensions into the 4 dimensions of spacetime can be carried out using models like the Kaluza-Klein model \cite{KK} and the Calabi-Yau model \cite{CY,yau}. Kaluza-Klein theory was proposed to unify electromagnetic theory and gravity in 5 dimensions. Supersymmetry (SUSY) can be accommodated in a Calabi-Yau manifold without symmetry breaking and hence can be considered as the best compactification model so far.  A paradigm shift in string theory came with the unification of these theories into a new supergravity M-theory with 11 dimensions \cite{HW,luka}.\\
\\
Efforts have been made to study inflation by incorporating supersymmetry into the general theory of relativity called supergravity (SUGRA). But they are not completely successful \cite{sugra}. Therefore, brane inflation with more than one supersymmetry (an extended $\mathcal{N}>1$ supersymmetric gauge theory where $\mathcal{N}$ is a collection of supercharges), where the matter content of the model is given by a $U(1)$ vector multiplet and a charged hypermultiplet \cite{all}, was introduced. The ground state of brane modes is considered the inflaton and is an open string with its two ends residing on different branes. The vacuum energy density drives inflation due to the initial separation of branes.   Inflation occurs through the interbrane interaction and can be described by the slowly rolling inflaton field. The resulting brane motion can bring the different branes together that were initially separated  \cite{dbrane,edi,braneint}. Inflation ends as soon as the branes approach and reach a critical distance, i.e., when the inflaton moves away from the unstable state and occupies the supersymmetric vacuum. Several attempts have been made to compute the tensor-to-scalar ratio from string theory motivated inflationary models. The corresponding tensor-to-scalar ratio can be compared with CMB estimates and can provide a platform to test the underlying inflationary model and hence the string theory \cite{j1,j2}. \\
\\
In standard cosmology, the inflationary energy scale ($\sim 10^{16}$ GeV) is much higher than the currently running Large Hadron Collider (LHC) experiments.
 Moreover, earlier LHC experiments were unsuccessful in providing any evidence for SUSY. This could be due to the extremely high energy scale of SUSY compared to the LHC. The misconception about the energy scale of SUSY can lead to a wrong conclusion about SUSY and related phenomena. This picture can be altered if we can get some evidence or signal from early universe phenomena such as superstring theory driven loop inflation through a suitable cosmological probe such as the cosmic microwave background (CMB).\\
 \\
Recently, the one loop inflation has gained much attention in exploring SUSY, brane inflation and Yukawa coupling. In the present work, we study the one loop corrected $D$-term inflation, which is a special case of a generic hybrid $P$-term brane inflation in a Calabi-Yau manifold in the context of type-IIB superstring theory. We show that the scalar spectral index and tensor-to-scalar ratio of one loop inflation are incompatible with the recent CMB estimates while using the Yukawa coupling from the Standard Model (SM) particle physics sector. In SM particle physics phenomenology, the well celebrated Yukawa coupling provides the strength of the interaction between fermions and bosons. Existing values \cite{atls,cms,arx} of the Yukawa coupling are estimated mostly from various SM particle interactions and such values are used to estimate the masses of quarks, leptons, etc. These coupling strengths can be challenged in the inflationary models stemming from supersymmetric potentials. The superpotential and the Yukawa couplings are determined by the underlying SUSY. Hence, we derive an alternative bound on the Yukawa coupling arising in the supersymmetric interactions from the cosmological sector. Further, we show that the tensor-to-scalar ratio of loop inflation with the Yukawa coupling deduced from cosmology is compatible with the recent CMB results and can be crucial in validating SUSY. Consequences of the Yukawa coupling obtained from the cosmological sector are also discussed.\\

\section{Loop inflation}
Recently, the loop inflation has received much attention in cosmology. The genesis of one loop inflationary model is from $\mathcal{N}=2$ hypersymmetric $P$-term inflation implemented in the D3/D7 brane that reduces to a supersymmetric $\mathcal{N}=1$ $D$-term inflation by modifying the flatness of the inflationary potential via one loop radiative correction \cite{enc,pdf,d3d7,dtrm,Nsusy,matsu}. 

\subsection{$P$-term inflation}
$P$-term inflation is usually defined in the framework of superconformal $SU(2,2|2)$ SUSY with the auxiliary fields of the vector multiplets $(0, \frac{1}{2}^2,1)$ implemented in D3/D7 branes, i.e, one end of the superstring lies on the D3 brane and the other end on the D7 brane \cite{pdf}. The inflation occurs in the slow roll regime where the D3 brane moves towards the D7 brane and ends with the symmetry breaking \cite{d3d7}. We assume the interaction of the inflaton field ($\phi$) with the positively and negatively charged superfields (respectively, $\varphi_+$ and $\varphi_-$)  in de-Sitter spacetime with a positive cosmological constant ($\Lambda$). 
The de-Sitter stage requires a soft symmetry breaking of $SU(2,2|2)$ to $\mathcal{N}=2$. This can be achieved by introducing a Fayet-Iliopoulos term (FI) $\xi^r $ ($r$=1, 2, 3),
\begin{align}
\xi &\equiv \sqrt{|\overrightarrow\xi|^2} =  \sqrt{\xi_+ \xi_- + \xi_3^2},\\
\xi_{\pm} &= \xi_1 \pm i \xi_2,
\end{align}
that can point in any direction. The global interaction potential for a generic extended supersymmetric $P$-term inflationary model can be written as
\begin{eqnarray}\label{this} 
V_{\mathcal{N}=2}^P & = y^2 \bigg( |\phi\varphi_+|^2 + |\phi\varphi_-|^2 + |\varphi_+\varphi_- - \frac{\xi_+}{2}|^2 \bigg) + \frac{\alpha^2}{2}(|\varphi_+|^2 - |\varphi_-|^2 - \xi_3)^2, 
\end{eqnarray}  
where $y$ is the holomorphic Yukawa coupling and $\alpha$ is the gauge coupling. The potential, Eq.(\ref{this}), corresponds to a ($\mathcal{N}=1$) supersymmetric model with the following choice of superpotential ($W$) and $D$-term 

\begin{align}
W &= y \phi (\varphi_+ \varphi_- - \frac{\xi_+}{2}),\\
D &= \frac{\alpha^2}{2}(|\varphi_+|^2 - |\varphi_-|^2 - \xi_3)^2.
\end{align}
The potential, Eq.(\ref{this}), has two minima, non-supersymmetric unstable vacuum  ($|\phi_c|^2 = \frac{\xi}{2}, \ \varphi_+ = \varphi_- =0$) through which the inflaton rolls slowly towards a supersymmetric global minimum ($\phi = 0, \  |\varphi_+|^2 - \ |\varphi_-|^2 - \xi_3=0 $, \ $|\varphi_+|^2 + \ |\varphi_-|^2 = \xi$) achieved through symmetry breaking, which is essential for the end of inflation. Here, $|\phi_c|$ represents the critical value of the inflaton field.  
Inflation requires a flat direction of the potential for which one can choose $\xi_1 =\xi_2 =0 $ and $\xi= \xi_3$. Also, during inflation the inflaton field $\phi$ dominates over the other moduli fields ($\varphi_+ = \varphi_- =0$) therefore the inflationary potential is
\begin{equation}
V_{inf} =  \frac{\alpha^2}{2} \xi^2.
\end{equation}
Inflation cannot be terminated as long as the inflationary potential remains flat. Therefore, to drive the inflationary potential towards its true vacuum, the flatness of the potential must be modified. This can be achieved by including loop corrections. Therefore, the effective potential of $P$-term inflation with loop corrections can be written as
\begin{equation}
V_{loop} =   \frac{\alpha^2\xi^2}{2} \bigg[ 1 + \frac{\alpha^2}{8\pi^2} \ln\frac{|\phi|^2}{|\phi_c|^2}  + f \frac{|\phi|^4}{2M_p^4}+ \mathcal{O}(\phi) \bigg] 
\end{equation}
where,  $ f = \frac{ \xi_1^2 + \xi_2^2 }{ \xi^2 }$  , $0<f<1$ and $\mathcal{O}(\phi)$ represents the higher order corrections. A generic $P$-term inflation is a combination of two supersymmetric inflationary models, $F$-term inflation with $\mathcal{N}=1$ chiral multiplet $(0, \frac{1}{2})$ and $D$-term inflation with $\mathcal{N}=1$ vector multiplet $(\frac{1}{2},1)$ respectively. But for inflation to happen smoothly, the massless $D$-term inflation must dominate over the $F$-term \cite{edi}. Whether $P$-term inflation proceeds through $D$-term or $F$-term is decided by the value of the FI term.  In the case of $D$-term model,  $f = \xi_+ = 0$, $|\varphi_+|
^2 = \xi_3$ and $\varphi_- = 0$ and for $F$-term $|\varphi_-|^2 = |\varphi_+|^2 = \xi$ . $f=1$ corresponds to $F$-term inflation in supergravity.

\subsection{$D$-term inflation}
In supergravity, the Yukawa coupling appearing in the superpotential is holomorphic ($y$), which is related to the physical Yukawa coupling ($\lambda$) by the relation $y = e^{K} \lambda$ \cite{holo,thesis}. Here the exponential factor containing the Kähler potential ($K$) poses some difficulties in implementing inflation (for details see \cite{sugra,dtrm, Einst, stew}). $D$-term inflation, being a locally supersymmetric model, escapes the contribution from $K$ and naturally avoids the problems faced in usual supergravity based models \cite{dtrm}. Hence, it is a suitable model to investigate the physical Yukawa coupling strength from supersymmetric interactions without any complexities. SUSY breaking is essential to maintain the flatness of the potential \cite{suzuki} so that the inflation proceeds in the slow roll regime. After the SUSY breaking, physical Yukawa couplings become much more relevant than the holomorphic Yukawa coupling. Therefore, in the rest of the text, we refer to the physical Yukawa coupling as Yukawa coupling.\\
\\
A hybrid hypersymmetric $P$-term inflation can be reduced to a supersymmetric $D$-term inflation by choosing $f=0$ under the special condition where the gauge coupling is related to the Yukawa coupling, $\alpha =\frac{\lambda}{\sqrt{2}}$ \cite{enc, dtrm, Nsusy}. $D$-term inflation is radiatively corrected upto one loop order \cite{dtrm,matsu}. Therefore, the global interaction potential and its one loop correction become
\begin{eqnarray}
V^D_{\mathcal{N}=1} &=& 2\alpha^2 (| \phi\varphi_-|^2 + | \phi\varphi_+|^2 +  | \varphi_{+}  \varphi_{-} |^{2}) + \frac{\alpha^2}{2} (| \varphi_{+} |^{2}  -| \varphi_{-} |^{2} + \xi)^2, \\
V_{one-loop} &=&   \frac{\alpha^2\xi^2}{2} \bigg[ 1 + \frac{\alpha^2}{8\pi^2} \ln\frac{|\phi|^2}{|\phi_c|^2} \bigg]. \label{eff}
\end{eqnarray}
Here, the critical value of the inflaton field $|\phi_c| = m_{pl}$. The resulting Einstein-Hilbert action on the brane is
\begin{equation}
S = \int d^4 x \ \sqrt{-g} \ \bigg[ \frac{m_{pl}^2R}{2} +  g^{\mu \nu} \partial_\mu \phi \partial_\nu \phi \\
+ V_{one-loop}\bigg].
\end{equation} 
The first and second slow-roll parameters for $V_{one-loop}$ in terms of $\alpha$ and the e-folding number ($N$) respectively can be expressed as
\begin{align}
\epsilon &= \frac{\alpha^2}{ \alpha^2 + 32  \pi^2 N}, \label{ep}\\ 
\eta &= \frac{-1}{4N\bigg(1+\frac{\alpha^2}{32 \pi^2 N}\bigg)}. \label{et}
\end{align}
The corresponding scalar spectral index and tensor-to-scalar ratio are
\begin{align}
n_s &= 1- \bigg(\frac{6\alpha^2 + 16\pi^2}{\alpha^2+32\pi^2N}\bigg),\label{cr1} \\ 
r &= \frac{16 \alpha^2}{ \alpha^2 + 32  \pi^2 N}. \label{cr2}
\end{align}
\begin{figure}[h]
\centering
\includegraphics[width=0.5\textwidth]{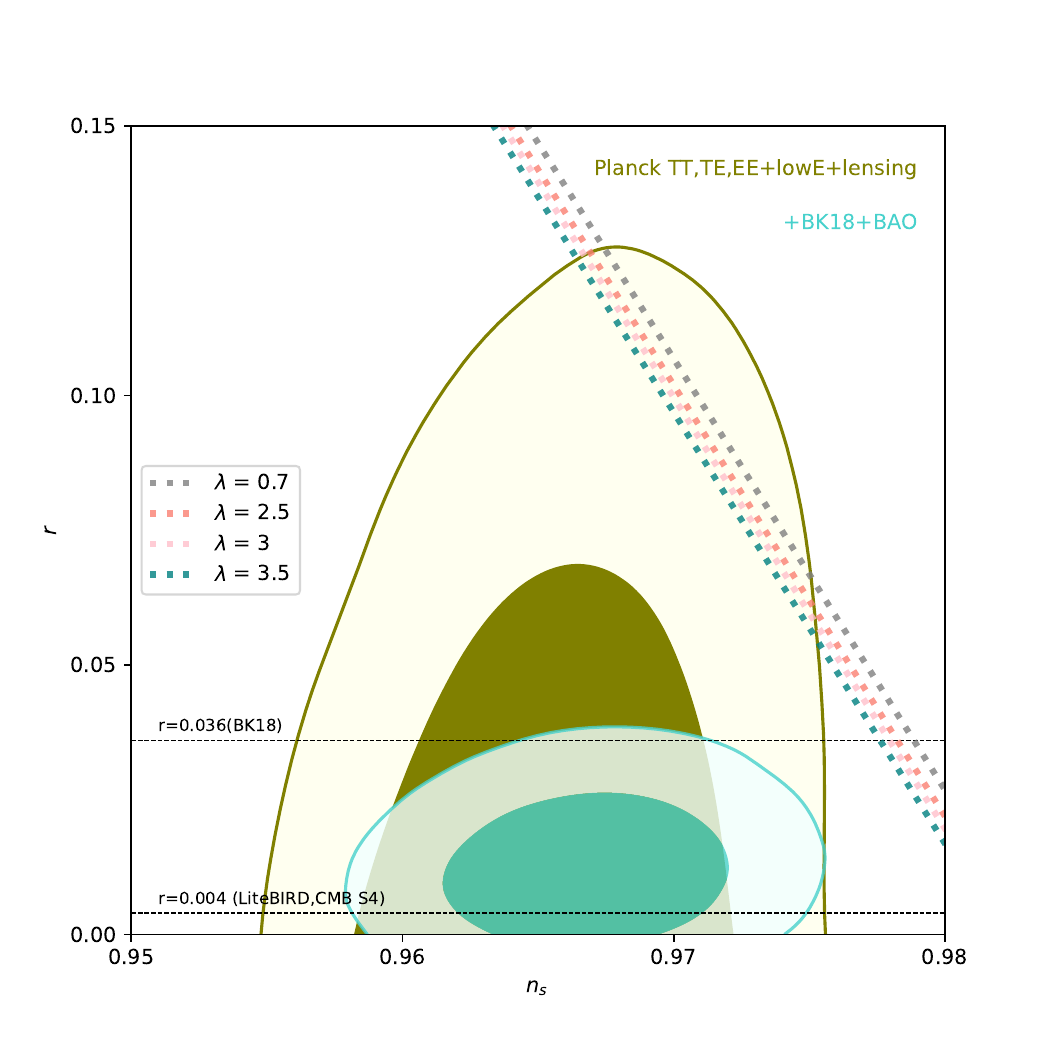}
\caption{\label{rvsns1} The tensor-to-scalar ratio $(r)$ of the one loop inflation for various values of scalar spectral index ($n_s$) and the SMYC  $0.7<\lambda<3.5$ with marginalized 95$\%$ and 68$\%$ CL  from the joint Planck TT,TE,EE+lowE+lensing and BK18+BAO.}
\end{figure}
\noindent The detailed derivation of Eqs. (\ref{ep}) and (\ref{et}) is provided in the appendix. We study the scalar spectral index and tensor-to-scalar ratio of one loop inflation with gauge coupling estimated from the presently available Yukawa coupling values from standard model particle physics \cite{atls,cms,arx}.
Next, we check the feasibility of this result for various parameters of the one loop inflation with the recent CMB estimates. From the comparative study (see Fig.\ref{rvsns1}), it can be observed that the tensor-to-scalar ratio of the loop inflation with the current bound from SMYC for the range $0.6<\lambda<3.5$ \cite{atls,cms,arx} is much higher than the upper limit derived from BK18 (0.036) \cite{par}, PR 4 + BK18 (0.032) \cite{find}, LiteBIRD \cite{hazumi} and CMB S4 (0.004, future projected estimate) \cite{s4}. Analysis of the variation of tensor-to-scalar ratio with the scalar spectral index shows that all $0.6<\lambda<3.5$ are out of the 95$\%$ and 68$\%$ CL of both Planck18 and BK18 bounds (see Fig.\ref{rvsns1}). Clearly, it can be noticed that for all  $0.6<\lambda<3.5$, the loop inflationary scenario is disfavoured based on the current and future projected CMB observations with the existing values of the Yukawa coupling constant from SM particle physics sector under consideration. This alarming situation prevents us from adopting SMYC directly into the cosmological sector. Since the loop inflation originates from the superstring theory, discarding the model can in turn challenge supersymmetry, string theory and quantum gravity. \\

\section{Yukawa coupling from CMB}
\begin{figure}[h]
\centering
\includegraphics[width=0.5\textwidth]{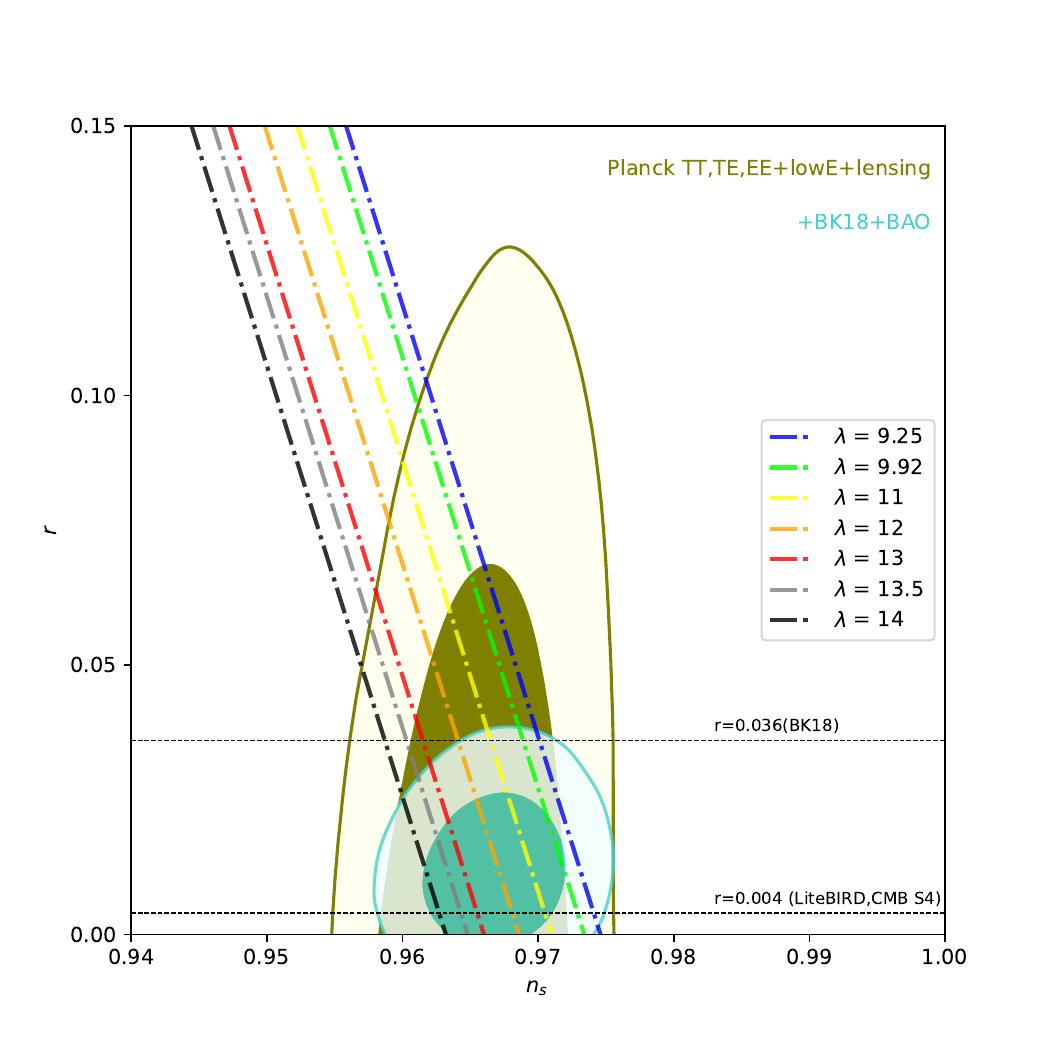}
\caption{\label{rvsns2} The tensor-to-scalar ratio $(r)$ of the one loop inflation for various values of scalar spectral index ($n_s$) and the CSYC $11<\lambda<14$ with marginalized 95$\%$ and 68$\%$ CL  from the joint Planck TT,TE,EE+lowE+lensing and BK18+BAO.}
\end{figure}

There is a necessity to investigate the values of Yukawa coupling beyond SM particle physics interactions, particularly suitable for the string motivated inflationary model. For this, the Yukawa interaction strength in the string driven cosmology needs to be derived. Hence, using Eq.(\ref{cr1}) and Eq.(\ref{cr2}), we estimate values of the Yukawa coupling constant for the one loop inflationary model from Planck18 and BK18. 
We study the feasibility of CSYC with the scalar spectral index and tensor-to-scalar ratio obtained from the joint Planck TT,TE,EE+lowE+lensing \cite{planck} and BK18+BAO \cite{par,updates} with the marginalized 95$\%$ and 68$\%$ CL (see Fig.\ref{rvsns2}). The analysis shows that $\lambda = 9.25$ falls within the 68$\%$ CL of Planck18 and  95$\%$ of BK18. But, all $\lambda < 9.25$ fall within the 95$\%$ CL of Planck18; however, they are outside the 95$\%$ and 68$\%$ CL of BK18 bounds and 68$\%$ CL  of Planck18. $\lambda = 9.92$ falls within the 68$\%$ CL of BK18 but it do not favor the future CMB forecasts. Similarly, all the values of $\lambda>14$ fall outside the 95$\%$ CL of both BK18 and Planck18, hence are not feasible CSYC for the one loop inflation. Therefore, to account for the current and future estimated tensor-to-scalar ratio, the situation enforces us to adopt a different range of the Yukawa coupling for the one loop inflation. This suggests an alternative model dependent bound on the Yukawa coupling applicable in the cosmological sector from the CMB, which is $11<\lambda<14$. \\

\noindent The one loop inflationary model originates from supersymmetric theory. Therefore, the Yukawa coupling strength associated with it can play a pivotal role in deciding its fate in the string theory based inflationary scenario. Hence, the right value of the Yukawa coupling must be used so that the model is compatible with various CMB observations, especially the tensor-to-scalar ratio. As we have mentioned, presently, values of the Yukawa coupling strength are estimated only from various SM particle physics sectors and exhibit a very wide range.  However, selecting the coupling strength from a pool of SM particle physics sector and using it in the loop inflation leads to inconsistency of its scalar spectral index and tensor-to-scalar ratio compared with the respective estimates from current CMB. This raises a concern about the adoptability of Yukawa coupling from the SM particle physics sector into the cosmological sector in a straightforward manner. This circumstance demands investigating suitable values of the Yukawa coupling constant for the viability of one loop inflation with various results of CMB. Therefore, we derived an alternate model dependent bound for Yukawa coupling from 68$\%$ CL of Planck18 and BK18 from CMB that is found applicable to cosmology. To the best of our knowledge, the authors are not aware of such a bound derived on Yukawa coupling from any cosmological sector so far. The present work is a key to overcoming the common misperception about the scale of SUSY in particle physics and cosmology. The newly derived constraint on the Yukawa coupling constant from CMB may play a crucial role in validating the inflationary model coming from supersymmetry. Yukawa coupling from string theory, supersymmetry, quantum gravity and extra dimensions can bridge the SM particle physics and cosmological sectors. 
\\
\section{Discussion and conclusion}
It is intriguing to note that SUSY is not established observationally to date, even at various high energy accelerator experiments such as the LHC, etc. Therefore, an alternate suggestion is to explore SUSY with an inflationary model originating from supersymmetry in the light of recent CMB observations. In superstring theory, the actual interaction strength is given by the physical Yukawa coupling, which is generally obtained by multiplying the holomorphic Yukawa coupling with a correction term related to the Kähler potential. Directly extracting the physical Yukawa coupling from the unknown geometry of the extra dimensional manifold is a tedious task. Therefore, the current work is much more convenient in estimating a new bound on the physical Yukawa coupling arising from superstring theory.\\
\\
Upon examining the superstring theory based loop inflation with the recent CMB observations, it is found that the model is disfavoured due to the adoption of SMYC in the associated fermionic and bosonic interactions. In SMYC, the Yukawa coupling is estimated from the high energy particle collider experiments such as ATLAS, CMS, etc. \cite{atls,cms,arx}. Another possibility is estimating the Yukawa coupling from cosmology by considering the one loop inflation through CMB.  We investigated the one loop inflation emerging from the superpotential and studied its tensor-to-scalar ratio and scalar spectral index with the 68$\%$  CL of both  Planck18 TT,TE,EE+lowE+lensing and BK18+BAO and demonstrated that the Yukawa coupling is possible beyond the SM particle physics sector. This constraint $11<\lambda<14$ from CMB can substantiate SUSY and its related phenomena. In short, we can say that the estimated Yukawa coupling from the cosmological sector and SM particle physics sector can complement to deepen the understanding so that the limitation of the LHC experiments can be overcome by looking at the cosmological phenomenology in estimating the Yukawa coupling for interactions taking place at energies higher than the LHC operating scale.
That may lead to interesting results which in turn complement understanding of particle physics, supersymmetry, string theory, quantum gravity and cosmology. Therefore, it can provide a platform to compare the values of Yukawa coupling strength arising from the cosmological sector with the SM particle physics sector. The difference in the values of the Yukawa coupling from the two distinct prominent sectors (SMYC and CSYC) arises because of the involved energy scales of interaction in the different sectors. Interestingly, SMYC estimates lie in 95$\%$ CL of ATLAS/CMS \cite{atls,cms,arx}  whereas CSYC falls in the 68$\%$ CL of Planck18 and BK18.
The outcomes of the present work may be useful in overcoming the strong criticism against the phenomenological realisation of string theory and its related ideas unambiguously. This may once again resurrect the wonderful idea of extra dimensions and superstring theory as a candidate for quantum gravity and can generate a renewed interest among the high energy and cosmology community. In that respect, our work takes a leap. \\
\\
An alternative bound on the Yukawa coupling indicates the existence of fermions with very high mass. The question is what could have happened to such fermions? One of the possibilities is that the high mass fermion decayed into other smaller particles. Another possibility is that the decayed fermion's density was considerably reduced because of inflation so that we don't see them in the present universe. At the same time a completely different aspect of it is that the production of such massive fermions is unlikely to happen which implies no signal for extra dimensions and superstring theory. The existence of an alternative bound for the Yukawa coupling from CMB therefore supports the same conclusion regarding the extra dimension and supersymmetry as that of the LHC experiment. Therefore, the result of the present work may also support the LHC conclusions. There will be many consequences if this bound is not acceptable, including the ruling out of loop inflation, which in turn negates supersymmetry. The study can be repeated with inflationary models arising from string theory that may support, widen or narrow the derived Yukawa coupling bounds. We hope that the current result can revive superstring and quantum gravity. Further, the bound on the Yukawa coupling derived from loop inflation may be re-examined with the upcoming CMB missions. A wide range of interactions and models have to be studied to derive a fundamental bound on the Yukawa coupling. The current work represents an initial step toward identifying a spectrum of permissible values from cosmology.

\section*{Appendix}
In this paper, we have considered the loop inflation with a potential,
\begin{eqnarray*} V_{one-loop} =  \frac{\alpha^2\xi^2}{2} \bigg[ 1 + \frac{\alpha^2}{8\pi^2} \ln\frac{|\phi|^2}{|\phi_c|^2} \bigg]. \end{eqnarray*}
However, it can be generalised to 
\begin{eqnarray*}V = M^4 \bigg[1+\beta \ \ln \bigg(\frac{\varphi}{m_{pl}}\bigg) \bigg], \end{eqnarray*} 
as given in \cite{enc} where, $M^4 = \frac{\alpha^2\xi^2}{2}$, $\beta = \frac{\alpha^2}{8\pi^2}$, critical value of the field $|\phi_c| = m_{pl} = 1$ in natural units and $|\phi|^2 = \varphi$. In this case, the first slow roll parameter can be obtained as,

\begin{eqnarray*} \epsilon = \frac{1}{2}(\frac{V'}{V})^2 = \frac{2\beta^2}{\varphi^2}\frac{1}{(1+\beta\ln\varphi)^2}. \end{eqnarray*} 
Solving this equation for $\varphi_{end}$ with the Lambert function as done in \cite{enc} yields $\alpha<0$. We do not prefer a negative value for the gauge coupling as it may lead to unphysical theories and hence look for $\alpha>0$ solutions. Since inflation takes place at sub-Planckian energy scales, $\frac{\varphi}{m_{pl}}<<1$. The Taylor series expansion of $\ln(x)$ centered at $a=1$ is,
\begin{eqnarray*}
 \ln(x)=\sum^{\infty}_{n=1} \frac{(-1)^{n-1}}{n}(x-1)^n.
  \end{eqnarray*} 
Therefore, $\ln(\varphi)\simeq \varphi-1$. As $\epsilon \rightarrow 1$, we have $\varphi_{end}^2[1+\beta(\varphi_{end}-1)]^2=2\beta^2.$ Neglecting higher orders in $\varphi_{end}$, we get $\varphi_{end} = \frac{\sqrt{2}\beta}{1-\beta}$. Next, we compute the number of e-folds of inflation,
\begin{eqnarray*}
N(\varphi)=-\int_{\varphi}^{\varphi_{end}} d\varphi \frac{V}{m_{pl}^2V'} = - \bigg[\frac{\varphi^2}{8}\bigg(\frac{1}{\beta}-1+\ln(\varphi)\bigg)\bigg]_{\varphi_N}^{\varphi_{end}}.
 \end{eqnarray*} 
 Following the same approximations for $\ln\varphi$ as before, we get,
\begin{eqnarray*}
N &=& \frac{\varphi_N^2 - \varphi_{end}^2}{8\beta}, \\
\varphi_N^2 &=& 8\beta N + \frac{2\beta^2}{(1-\beta)^2}\simeq 8\beta N+2\beta^2.
\end{eqnarray*} 
 Using $\varphi_N$, we express the first slow roll parameter in terms of N,
 \begin{eqnarray*}
 \epsilon =  \frac{2\beta^2}{(8\beta N+2\beta^2)}\frac{1}{(1+\beta\ln\varphi)^2}.
 \end{eqnarray*} 
$\varphi <<1$ and $\beta<<1$ implies $1+\beta \ln\varphi \simeq1$. Therefore,
 \begin{eqnarray*}
\epsilon = \frac{\beta}{4N+\beta} .
 \end{eqnarray*} 
This expression is similar to the $\epsilon_1^*\simeq \frac{\beta}{4N}$ discussed in \cite{enc}.
Substituting for $\beta$, we obtain,
 \begin{eqnarray*}
\epsilon = \frac{\alpha^2}{32\pi^2N+\alpha^2} .
 \end{eqnarray*} 
Similarly, we obtain the second slow roll parameter, 
\begin{eqnarray*}
\eta &=  m_{pl}^2\bigg(\frac{V''}{V}\bigg) = \frac{\frac{-2\beta}{\varphi^2}}{1+\beta\ln\varphi} \simeq \frac{-2\beta}{\varphi^2} .
\end{eqnarray*}
Substituting $\varphi_N$,  the second slow roll parameter can be rewritten in terms of $N$ as, 
\begin{eqnarray*}
\eta = \frac{-2\beta}{ 8N\beta+2\beta^2} = \frac{-1}{4N(1+\frac{\alpha^2}{32\pi^2N})} \simeq \frac{-1}{4N}.
\end{eqnarray*}

\section*{Acknowledgement}
AB acknowledges the financial support of Prime Minister’s Research Fellowship (PMRF ID : 3702550) provided by the Ministry of Education, Government of India.

\end{document}